\documentclass[aps,pra,reprint,showpacs]{revtex4-1}

\usepackage{graphicx}
\usepackage{amsfonts}
\usepackage{amssymb}
\usepackage{amsmath}
\usepackage{bm}
\newcommand{\be}{\begin{equation}}
\newcommand{\ee}{\end{equation}}
\newcommand{\bea}{\begin{eqnarray}}
\newcommand{\eea}{\end{eqnarray}}

\begin{document}

\title{Anisotropic sub-Doppler laser cooling in dysprosium magneto-optical traps}

\author{Seo Ho Youn}
\author{Mingwu Lu}
\author{Benjamin L. Lev}
\affiliation{Department of Physics, University of Illinois at Urbana-Champaign, Urbana, IL 61801-3080 USA}

\begin{abstract}
Magneto-optical traps (MOTs) of Er and Dy have recently been shown to exhibit populationwide sub-Doppler cooling due to their near degeneracy of excited- and ground-state Land\'{e} $g$ factors.  We discuss here an additional, unusual intra-MOT sub-Doppler cooling mechanism that appears when the total Dy MOT cooling laser intensity and magnetic quadrupole gradient increase beyond critical values.  Specifically, anisotropically sub-Doppler-cooled cores appear, and their orientation with respect to the quadrupole axis flips at a critical ratio of the MOT laser intensity along the quadrupole axis versus that in the plane of symmetry.  This phenomenon can be traced to a loss of the velocity-selective resonance at zero velocity in the cooling force along directions in which the atomic polarization is oriented by the quadrupole field.  We present data characterizing this anisotropic laser cooling phenomenon and discuss a qualitative model for its origin based on the extraordinarily large Dy magnetic moment and Dy's near degenerate $g$ factors.
\end{abstract}
\date{\today}
\maketitle

\section{Introduction}
Ultracold dipolar gases are emerging as an important tool for exploring strongly correlated physics in a system with easily and rapidly tunable parameters.  Highly magnetic atoms provide the ability to realize strongly correlated systems in the absence of polarizing fields, which will be important for exploring, e.g., quantum liquid crystal physics~\cite{Fregoso:2009,*Fregoso:2009b,*fregoso:2010}.  Additionally, degenerate fermionic and bosonic dipolar gases are expected to exhibit a wide range of interesting quantum phases and dynamical properties~\cite{PfauReview09,*Pu09,*Bohn10,*Machida10}.  Dysprosium, the most magnetic laser cooled and trapped gas~\cite{Lu2010,Youn2010a}, is especially exciting since its $\mu=10$~$\mu_{B}$ magnetic dipole moment---coupled with its large mass and abundance of bosonic and fermionic isotopes---is sufficiently strong to allow wide exploration of ultracold dipolar physics~\cite{Lu2010,Youn2010a,Yi:2007}.

Sub-Doppler cooling~\cite{Dalibard89,*Chu89} in an optical molasses~\cite{MetcalfBook99} subsequent to MOT capture is a crucial step toward creating degenerate gases in, e.g., alkali atomic systems such as Rb and Cs.  Sub-Doppler mechanisms fail to cool the majority of atoms~\cite{Salomon94} inside these MOTs because the ground state $g_{g}$ factor is substantially different from the excited state's (for $^{87}$Rb, $\Delta g/g_{g}=34$\%)~\cite{Phillips92,*Ertmer92}; population-wide $\sigma^{+}$-$\sigma^{-}$ sub-Doppler cooling breaks down in the longitudinal magnetic fields of a typical MOT.  Notable exceptions~\footnote{The $^{87}$Sr system, which has a negligible $g_{g}$, also exhibits intra-MOT sub-Doppler cooling~\cite{YeSr03}.} are the highly magnetic lanthanides Er (7 $\mu_{B}$) and Dy, whose repumper-less MOTs~\cite{Mcclelland:2006,Lu2010} operate on optical transitions whose ground and excited state $g$ factors are the same to within 0.3\% (1.7\%) for Er (Dy)~\cite{Martin:1978}.  At low MOT beam intensity, both the Er and Dy MOTs exhibit {\it{in-situ}}, population-wide sub-Doppler cooling~\cite{Berglund:2007,Youn2010a}.  The minimum temperatures $>$$200$ $\mu$K are, however, more than 10 times larger than typical Rb and Cs optical molasses sub-Doppler cooled temperatures, but are similar to sub-Doppler temperatures in the $^{87}$Sr MOT~\cite{YeSr03}.  
\begin{figure}[t]
\includegraphics[width=0.3\textwidth]{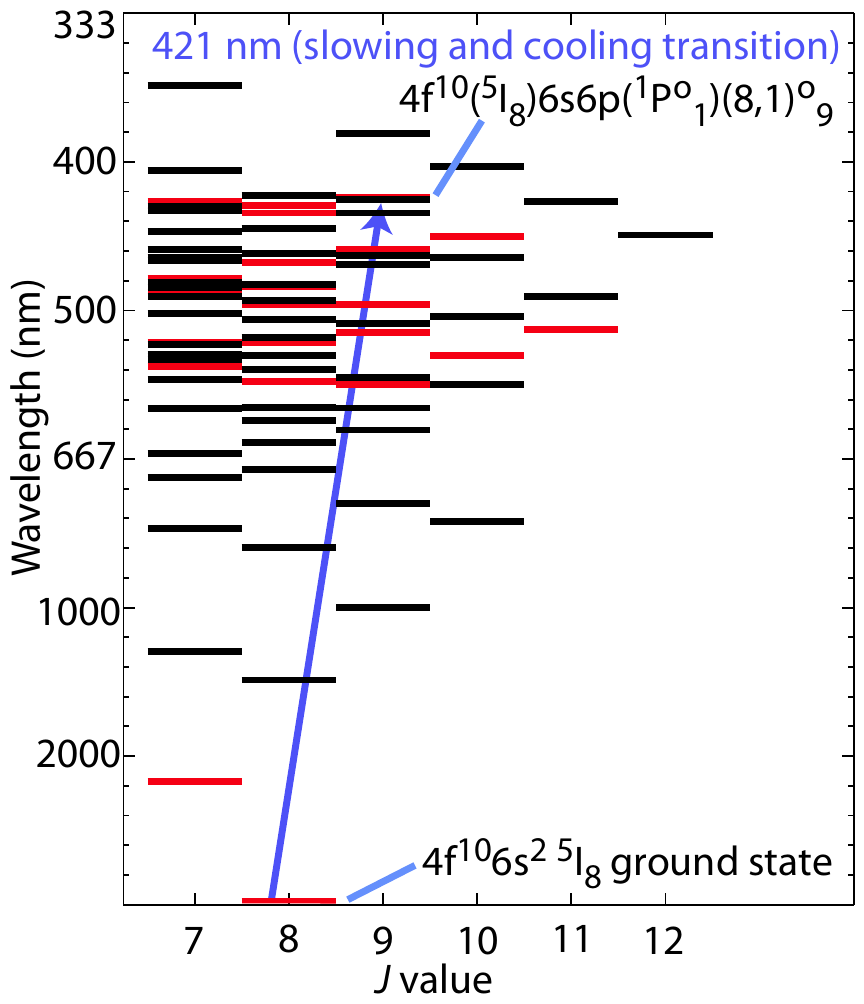}
\caption{(color online). $^{{164}}$Dy energy level structure~\cite{Martin:1978}.  The MOT and Zeeman slower employ the strongest laser cooling transition between the even parity (red) ground state and the odd (black) excited state at 421 nm. For this transition, the $g$ factor of the ground (excited) state is 1.24 (1.22) and the Doppler cooling limit is $T^{0}_{D}=768$ $\mu$K.  $J$ is the total electronic angular momentum.} 
\label{fig:dy_levels}
\end{figure}

We describe here a novel sub-Doppler cooling mechanism in the highly magnetic Dy MOT system that we first reported in Refs.~\cite{Lu2010,Youn2010a}.  We expand on these observations with additional measurements and offer a plausible explanation for this phenomenon based on the theory of velocity selective resonances (VSR) in a highly magnetic gas~\cite{Metcalf91,vanderStraten93,Lee02}.  While this mechanism only cools a small fraction of the atoms to $\sim$10 $\mu$K ultracold temperatures, its existence is a first example of novel behavior arising from laser cooling the most magnetic atom.  

\begin{figure}[!t]
\includegraphics[width=0.48\textwidth]{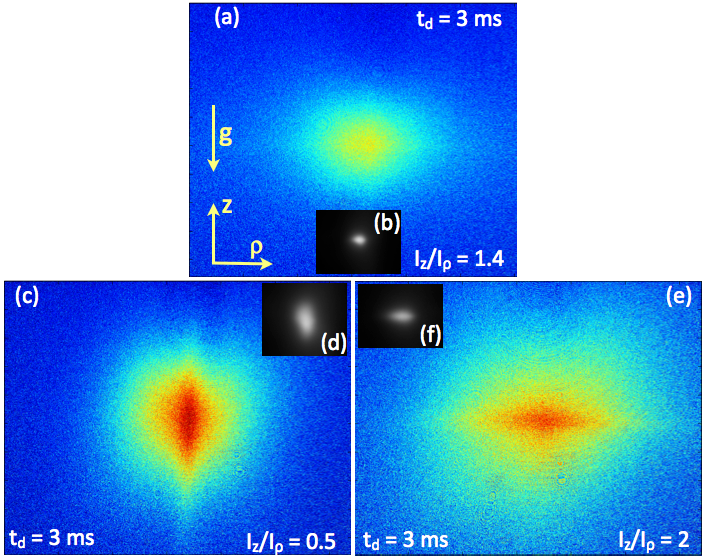}
\caption{\label{stripeimage} (color online).  (a,c,e) Time-of-flight expansions and (b,d,f) {\it{in-situ}} images of the three striped MOT regimes exhibiting anisotropically sub-Doppler cooled cores.  (a,b) Symmetric/cross MOT with $I_{z}/I_{\rho} =1.4$.  Notice the faint vertical and horizontal stripes crossing at the MOT center.  (c,d) Prolate striped MOT with $I_{z}/ I_{\rho} = 0.5$.  (e,f) Oblate striped MOT with $I_{z}/I_{\rho}=2$. }
\end{figure}
\section{Dy MOT regimes}
We choose to describe here the Dy MOT system in which all pairs of MOT beams are well-aligned in mutual orthogonality and retroflection~\footnote{Boundaries between the striped MOT regimes are blurred with substantial beam misalignment.}.  All data are taken with the experimental apparatus for Dy MOT production described in Ref.~\cite{Youn2010a}.  Briefly, a high-temperature oven along with transverse laser cooling produce a collimated atomic beam.  A Zeeman slower, operating with 1~W of laser power on the 421-nm cycling transition, decelerates the beam before MOT capture on the 421-nm line  (see Fig.~\ref{fig:dy_levels}).  No repumper is necessary since the highly magnetic atoms remain confined in the MOT's magnetic quadrupole trap as population decays through metastable states~\cite{Lu2010,Youn2010a}.  We focus here on the bosonic isotope $^{164}$Dy with zero nuclear spin $I$, but similar results were observed in the fermionic $^{163}$Dy MOT ($I=5/2$). 

The Dy MOT can be formed in two classes of operation, striped and stripeless, and the striped MOT can be further classified into three regimes.  The stripeless Dy MOT is similar to the Er MOT reported in Ref.~\cite{Berglund:2007}, in that the majority of atoms are sub-Doppler cooled to $\sim$200 $\mu$K.  These MOTs are obtained at low intensity $\bar{I}=I_{t}/I_{s}\alt0.17$, where $I_{t}=I_{z}+2I_{\rho}$ is the total MOT intensity (one beam pair in $\hat{z}$ and two in $\hat{\rho}$) and $I_{s}$ is the saturation intensity in the MOT; $I_{s} = 2.7\times58$ mW/cm$^{2}$~\cite{Youn2010a}.  Moreover, the stripeless $^{164}$Dy MOT forms at low MOT magnetic quadrupole field $\nabla_{z} B\alt12$ G/cm, where $\nabla_{z} B\approx2\nabla_{\rho} B$ and $\hat{z}$ and $\hat{\rho}$ are the directions along the quadrupole's axis and plane of symmetry, respectively.  

The temperature and density distributions in the three striped MOT regimes differ greatly from the stripeless MOT in time-of-flight.  The temperature distribution is anisotropic at large $\bar{I}$ and $\nabla_{z} B$, and Fig.~\ref{stripeimage} shows characteristic time-of-flight expansions of the striped Dy MOT.  In these images, MOT beams and the quadrupole field are extinguished at $t_{d}=0$, and the {\it{in-situ}} images are taken at $t_{d}<0$.   We observe in time-of-flight a low-population core surrounded by a more populous and hotter outer shell, and we designate the group of outer shell atoms as the majority while the inner core as the minority.  
Parametrically driving the MOT could reveal the minority component's {\it{in-situ}} size~\cite{Jhe:2004}.

\begin{figure}[t]
\includegraphics[width=0.49\textwidth]{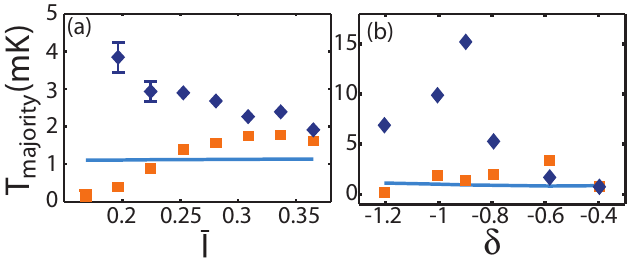}
\caption{\label{MA}(color online).  Temperature characterization of the prolate striped Dy MOT [Fig.~\ref{stripeimage}(c)] versus normalized intensity $\bar{I}$ and detuning $\delta$.  The orange squares (blue diamonds) are the $T_{\text{majority}}$ along $\hat{\rho}$ ($\hat{z}$).  Light blue curves show the Doppler cooling limit~\cite{MetcalfBook99} for the MOTs' parameters:  (a) $\delta = -1.2$, $\nabla_{z}\text{B}=20$ G/cm; (b) $\bar{I} = 0.2$, $\nabla_{z}\text{B}=20$ G/cm.}
\end{figure}

\begin{figure}[t]
\includegraphics[width=0.49\textwidth]{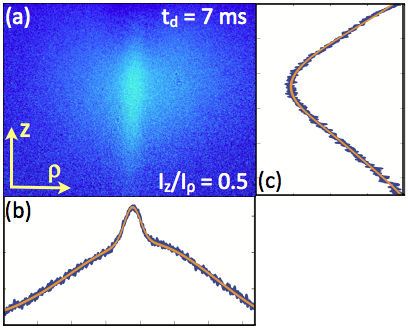}
\caption{\label{stripeimagefit}(color online).  Prolate striped MOT time-of-flight expansion at $t_{d}=7$ ms. (a) Double-Gaussian fit, orange line, along a $\hat{z}$ intensity integration.   (b) Intensity integration along $\hat{\rho}$ is consistent with a single Gaussian fit. }
\end{figure}

Typical high-population MOTs are formed with $\bar{I}=0.2$ and $\delta=\Delta/\Gamma=-1.2$, where $\Delta$ is the detuning from the $\Gamma = 2\pi\cdot32$ MHz, 421-nm transition.  These typical MOT parameters correspond to $T_{D}$'s of approximately 1 mK.  At large MOT intensity, the temperature of the majority component $T_{\text{majority}}$ is consistent with $T_{D}$, as shown in the $\hat{\rho}$ (orange square) data of Fig.~\ref{MA} and as discussed in detail in Ref.~\cite{Youn2010a}.    The $T_{\text{majority}}$ data in $\hat{\rho}$  are derived from the broader of the two Gaussians employed to fit intensity versus $\rho$ [see Fig.~\ref{stripeimagefit}(b)].   Temperature data for $\hat{z}$ (blue diamond) are derived from single Gaussian fits to intensity versus $z$ [see Fig.~\ref{stripeimagefit}(c)].  

The larger $\hat{z}$ temperature data are a result of the convolution of the Doppler-cooled majority temperatures with the minority velocity distribution along $\hat{z}$ (described in Sec.~\ref{proof}).  We note that $T_{\text{majority}}$ data in $\hat{z}$ below $\bar{I}\approx 0.17$ are roughly equal to the $\hat{\rho}$ data and are not shown.  This low-intensity, sub-Doppler cooled region is also the region in which stripes cease to appear in time-of-flight expansions:  In the Dy system, the stripeless regime is the low-intensity limit of the striped regime.  We note that the sub-Doppler-cooled Er MOT, which did not exhibit stripes, was also operated below $\bar{I}=0.17$ in the data of Ref.~\cite{Berglund:2007}.

\begin{figure}[t]
\includegraphics[width=0.49\textwidth]{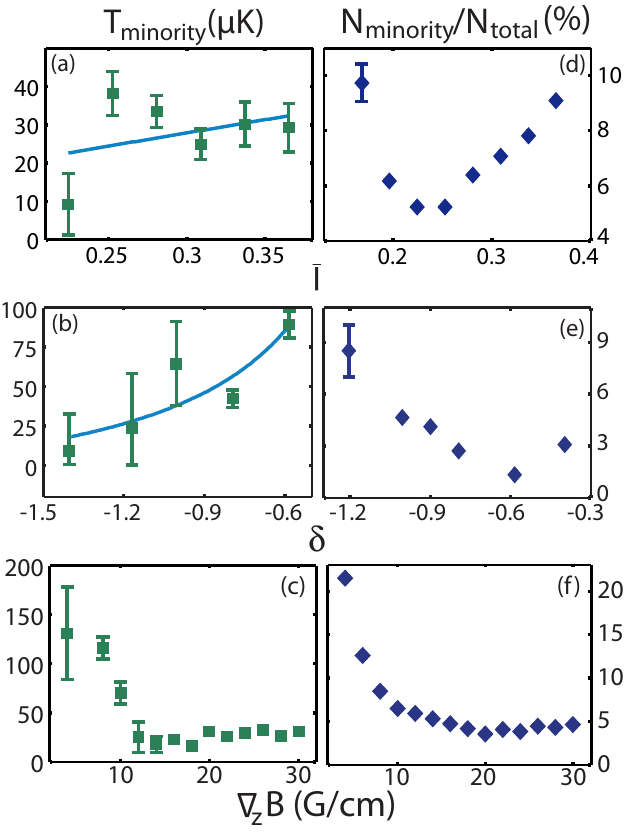}
\caption{\label{MI}(color online).   $T_\text{minority}$ in $\hat{\rho}$ of the prolate striped MOT $I_{z}/I_{\rho}=0.5$  [Fig.~\ref{stripeimage}(c)].  Blue curves in (a, b) are fits of the data to the sub-Doppler cooling expression Eq.~\ref{SubDop}.  (d--f) Fractional population of the minority component. (a, d) $\delta = -1.2$, $\nabla_{z}\text{B}=20$ G/cm; (b, e) $\bar{I} = 0.2$, $\nabla_{z}\text{B}=20$ G/cm; (c, f) $\bar{I}=0.2$, $\delta=-1$.}
\end{figure}

\begin{figure}[t]
\includegraphics[width=0.49\textwidth]{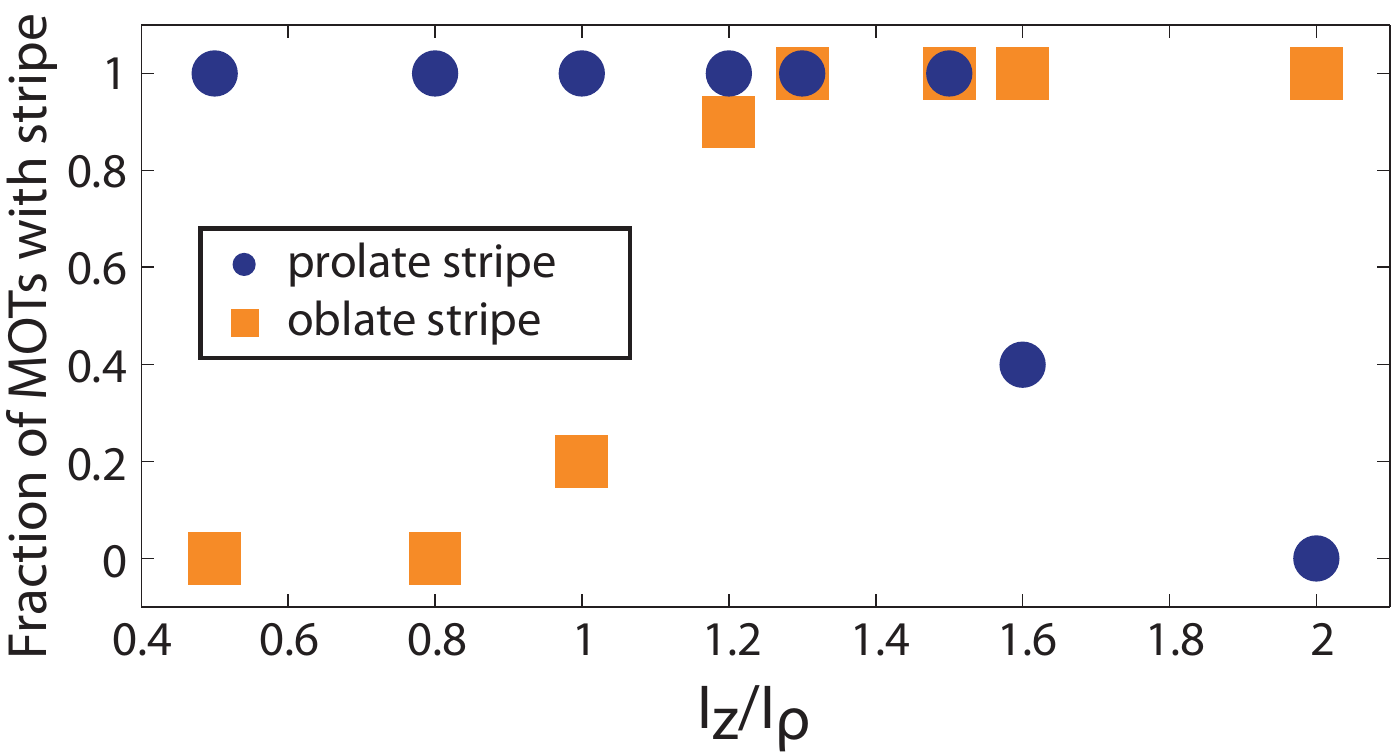}
\caption{\label{VsPz}(color online).  Visibility of stripe regimes---prolate, symmetric/cross, and oblate---versus the ratio of MOT beam power $I_{z}/I_{\rho}$.  Data are taken for $\bar{I}=0.22(0.01)$, $\delta=-1$, and $\nabla_{z}\text{B}=20$ G/cm.  All stripes vanish below $\bar{I}_{c}=0.17$.  Each point is the average of 10 MOT realizations with $t_{d}=6.5$ ms.}
\end{figure}

\section{Quantitative characterization of anisotropic sub-Doppler cooling}

The minority component assumes either a prolate, symmetric/cross, or oblate shape in time-of-flight depending on whether the ratio of intensity in the MOT beams $I_{z}/I_{\rho}$ is $<$1.4, $\sim$1.4, or $>$1.4 (see Fig.~\ref{stripeimage}).  MOT populations are largest in the prolate, $I_{z}/I_{\rho}<1.4$ regime, whose minority component temperature $T_{\text{minority}}$ and ratio of minority atoms to total MOT population $N_{\text{minority}}/N_{\text{total}}$ is shown in Fig.~\ref{MI}~\footnote{Careful characterization of minority population and temperature for MOTs with $I_{z}/I_{\rho}>1$ was inhibited by increased MOT population instability in these regimes.}. 

$T_{\text{minority}}$ and $N_{\text{minority}}$ are extracted from a double-Gaussian fit to time-of-fight data such as that presented in Fig.~\ref{stripeimagefit}.  The temperature of the minority component is anisoptropic, leading to the vertical stripe structure in Figs.~\ref{stripeimage}(c) and~\ref{stripeimagefit}(a). The temperature in $\hat{z}$ is hotter than the majority atoms [see the $\hat{z}$ (blue diamond) data of Fig.~\ref{MA}], while the temperature in $\hat{\rho}$ is sub-Doppler cooled to temperatures as low as 10 $\mu$K, a factor of 10 less than the sub-Doppler temperatures observed in the stripeless regime.  The $\hat{\rho}$ temperature of the minority component in the prolate MOT is well described by a 1D characteristic sub-Doppler scaling law~\cite{Salomon94,YeSr03}:
\be\label{SubDop}
T_{\rho}=T_{0}+C_{\sigma^{+}\sigma^{-}}\frac{\hbar\Gamma\bar{I}}{2k_{B}|\delta|}
\ee
where $[C_{\sigma^{+}\sigma^{-}}]=[0.1(0.09), 0.55(0.20)]$ in Figs.~\ref{MI}(a) and (b), respectively.  These values of $[C_{\sigma^{+}\sigma^{-}}]$ are consistent with those of the Er MOT~\cite{Berglund:2007}, but smaller than in the $^{87}$Sr MOT~\cite{YeSr03}.  However, $T_{0}$ is $<$10 $\mu$K in both fits and is $\sim$$10\times$ lower than in the Er, Sr, and stripeless Dy MOTs.  This 2D, anisotropic sub-Doppler cooling---in the case of the oblate striped MOT, 1D~\footnote{We observe, by imaging in the $\rho$-plane, that the oblate stripe is indeed azimuthally symmetric.}---is a much more effective cooling mechanism than the one found in the stripeless sub-Doppler-cooled Er and Dy MOTs.  

\section{Transition between prolate and oblate striped MOTs}

We explore the transition between the prolate and oblate striped Dy MOT in Fig.~\ref{VsPz}.  As the MOT's $I_{z}/I_{\rho}$ ratio is tuned from 0.5 to 2, we note the occurrence---defined by visibility of the double-Gaussian structure above image noise---of stripes in a series of 10 realizations. The fraction of these images exhibiting a stripe structure is noted in Fig.~\ref{VsPz}.    Below $I_{z}/I_{\rho}=1$, the prolate stripe is always observed, while the oblate stripe is always observed above $I_{z}/I_{\rho} = 2$.    There is a smooth transition between the regimes in which successive MOT realizations may exhibit either a prolate or oblate stripe, and we can offer no explanation for either the smoothness of the transitions or their slight asymmetry.     

Between these two regimes there is a third:  We most often observe a spherically symmetric minority component in the range of $I_{z}/I_{\rho}$ from 1.3 to 1.5.  This is the regime of the power balanced and aligned MOT discussed in Ref.~\cite{Lu2010} and is less populous, though more dense, than the prolate regime.  A prolate--oblate cross is observed [see Fig.~\ref{stripeimage}(a)] in the region around the critical ratio $I^{c}_{r}=I_{z}/I_{\rho}=1.4$,  though not reproducibly, and the spherically symmetric core is more prevalent.  

No stripes are observed below the critical intensity $\bar{I}_{c}<0.17$, which coincides with the appearance of majority component (population-wide) sub-Doppler cooling [see Fig.~\ref{MA}].  The prolate stripe is observed to vanish below a critical $\nabla_{z}\text{B}_{c}\approx12$ G/cm, as shown in Fig.~\ref{MI}(c).  The minority component temperature and stripe population fraction rise below  $\nabla_{z}\text{B}_{c}$---which may be interpreted as a blending of the minority and majority components---until the MOT is firmly in the stripeless, isotropically sub-Doppler-cooled regime.   

\section{Velocity selective resonance picture of anisotropic sub-Doppler cooling}\label{VSR}

This section presents a qualitative explanation of the anisotropic sub-Doppler cooling mechanism.  Two properties of Dy are crucial to this explication: its large magnetic moment and the near equal ground and excited state $g$ factors ($\Delta g \approx 0$) of the cooling transition.  The latter is important for the following reason.  In standard $\sigma^{+}$-$\sigma^{-}$ sub-Doppler cooling, the linear optical polarization serves as the quantization axis since the optical pumping rate $\gamma_{p}$ is typically greater than the Larmor precession rate $\omega_{L}=\mu \nabla B r/\hbar$, where $r$ is the distance of the atom from the magnetic quadrupole's center.   In low magnetic fields, ground state population imbalance due to non-adiabatic following of the quantization axis induces the differential scattering of $\sigma$-light, which then leads to sub-Doppler cooling to zero mean velocity~\cite{Dalibard89,MetcalfBook99}.  However, in large longitudinal magnetic fields, a non-zero $\Delta g$ results in an ``unlocking'' of the Doppler and sub-Doppler-cooling mechanisms, and sample temperatures rise to the Doppler-cooling limit. (Hence the failure of population-wide sub-Doppler cooling in, e.g., Rb MOTs and the need for field nulling to the mG level when sub-Doppler cooling in Rb optical molasseses.)  By contrast, Er and Dy MOTs remain sub-Doppler-cooled~\cite{Berglund:2007,Youn2010a} despite the large intra-MOT magnetic fields because $\Delta g\approx0$ in both.  This explains the existence of intra-MOT sub-Doppler cooling in Er and Dy, but not the origin of the anisotropic regime.

The anisotropic sub-Doppler-cooling mechanism may be qualitatively understood in the velocity-selective resonance (VSR) picture of 1D sub-Doppler cooling~\cite{Metcalf91} when augmented to account for large magnetic fields, or equivalently, large magnetic moments~\footnote{A review of such a treatment is beyond the scope of the current work.  However, Refs.~\cite{vanderStraten93,Lee02} provide detailed analytical and numerical calculations of the force and diffusion felt by atoms in various relative orientations of (a large) magnetic field and $\sigma^{+}$-$\sigma^{-}$ light; see Figs.~4 and 5 and Table 1 of Ref.~\cite{vanderStraten93} and Fig.~4 of~\cite{Lee02} for force versus velocity plots and additional details of the VSR cases invoked in Sec.~\ref{VSR}}.  In this picture, most types of sub-Doppler cooling may be understood as arising from the momentum transfer by a coherent two-photon process between ground state sub-levels (labeled by $m$)~\cite{Metcalf91}.  These coherent Raman transitions can occur when the difference in Doppler-shifted frequencies of counterpropagating light beams seen by an atom equals the Zeeman shift between levels separated by $\Delta m = n$.  Specifically, VSRs occur when $\delta\bm{k}\cdot \bm{v} = n \mu_{B} B$, where $\bm{v}$ is the atom's velocity and $\delta\bm{k}=2k$ is the difference in the wavevectors of the $\sigma^{+}$-$\sigma^{-}$ light.  The allowed $n$'s depend on the relative orientation of optical and magnetic field vectors, and it is this selection rule that is the origin of the anisotropic sub-Doppler cooling in the large field (large magnetic moment) regime.

To see how this selection rule leads to anisotropic cooling, we first note that the atoms' polarization aligns along the local magnetic field in the presence of a large magnetic field (or large moment), defined as $\gamma_{p}\ll\omega_{L}\ll\Gamma$; in this large Zeeman energy regime, the magnetic field now serves as the quantization axis, not the optical polarization~\cite{vanderStraten93}.  Now we examine the case in which the magnetic field is aligned parallel to the $\bm{k}$'s of a pair of $\sigma^{+}$-$\sigma^{-}$ beams.  Along the spatial directions defined by the $\bm{k}$'s of the cooling light, the zero-mean-velocity VSR in the cooling force disappears ($n\neq0$) even though finite-velocity VSRs remain ($n\geq2$ are allowed).  This is because there is no component of the polarization along the magnetic field (azimuthal symmetry about $\bm{k}$ is preserved), and thus no $\pi$ transitions are allowed, which are necessary for $n=0$ and $n=1$ VSRs.  In this case, sub-Doppler cooling to zero velocity along directions close to $\bm{k}$ is not possible when $\bm{B}\parallel \bm{k}$.  

However, not all sub-Doppler cooling is suppressed.  A zero-velocity VSR remains in the cooling force along directions in which the local magnetic field is perpendicular to the cooling light's $\bm{k}$'s (for $\bm{B}\perp \bm{k}$ all $n$ are allowed).  This spatial modulation of allowable VSRs ($n$'s) is the origin of the anisotropic nature of sub-Doppler cooling observed in the highly magnetic Dy system. 
\begin{figure}[t]
\includegraphics[width=0.49\textwidth]{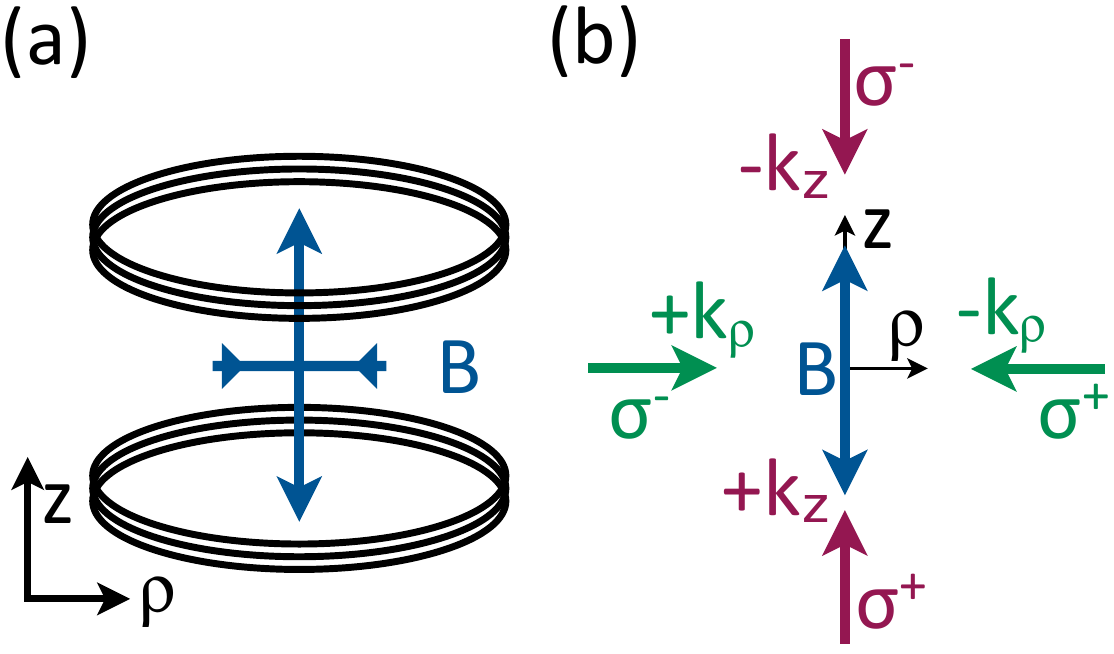}
\caption{\label{SD}(color online).  (a) Orientation of the MOT's magnetic quadrupole field.  (b) Illustration of the anisotropic cooling in the prolate MOT including the MOT magnetic ($B$) field orientation, cooling beam wavevectors $\bm{k}$, and polarizations $\sigma^{\pm}$ near the $z$-axis.}
\end{figure}

In summary, $\sigma^{+}$-$\sigma^{-}$ sub-Doppler cooling is preserved (fails) along directions perpendicular (parallel) to the local magnetic field in systems with large magnetic fields or large magnetic moments.  Figure~\ref{SD}(b) illustrates this idea for the prolate striped MOT:  sub-Doppler cooling occurs along $\hat{\rho}$ due to the green $\bm{k}_{\rho}$ $\sigma^{+}$-$\sigma^{-}$ beams, but not along $\hat{z}$ due to atomic polarization orientation parallel to the magnetic field.  The red $\bm{k}_{z}$ $\sigma^{+}$-$\sigma^{-}$ beams only induce VSRs with non-zero velocity (e.g., $n\geq2$) in the cooling force, and these beams cause the atoms to accelerate along $\hat{z}$ due to the magnetic quadrupole field gradient.  The analogous mechanism holds for the oblate MOT when the magnetic field (quantization axis) is along $\hat{\rho}$ instead.  

Prolate and oblate stripes oriented with respect to the quadrupole axis dominate because the only regions wherein the magnetic field is parallel to the $\bm{k}$-vectors are those in the plane or along the axis of symmetry of the quadrupole field.  The regions' small relative volume is the origin of $N_{\text{minority}}/N_{\text{total}}<10$\%.  We note that this $\sigma^{+}$-$\sigma^{-}$ mechanism is distinct from $\sigma^{+}$-$\sigma^{+}$ magnetically induced laser cooling~\cite{MetcalfBook99}. 

\section{Anisotropic sub-Doppler cooling in the Dy system}\label{proof}

Although this description does not predict the prolate-to-oblate transition, our quantitative measurements in the prolate regime are consistent with this picture.  Specifically, when $I_{z}/I_{\rho}<I_{r}^{c}$, $\bar{I}>\bar{I}_{c}$, and $\nabla_{z}\text{B}>\nabla_{z}\text{B}_{c}$ such that a prolate striped MOT forms, we observe higher-than-Doppler-limited temperatures along the $\hat{z}$ direction, which lies parallel to the local magnetic field (Fig.~\ref{MA}).  As mentioned in the previous section, this is because along the $\bm{B}\parallel k$ direction the atoms accelerate as they move away from the trap center due to ever increasing magnetic fields and VSR velocities.  An estimate of these resonant velocities---based on $v = \omega_{z}/k$~\cite{Metcalf91} in the 20 G/cm gradient field along $\hat{z}$---are $\geq T_{D}$, and are thus consistent with the absence of any sub-Doppler cooling mechanism along this direction.  Specifically, outside a core of radius 30 $\mu$m---roughly the zone outside which anisotropic sub-Doppler cooling occurs; see below---the velocities are $>$$0.35$ m/s, resulting in 1D temperatures of $>$$2.5$ mK.  Such temperatures are larger than the $T_{D}\approx1$ mK limit for the typical Dy MOT parameters (see blue curves in Fig.~\ref{MA}).  

Anisotropic sub-Doppler cooling is possible for the Dy MOT on the 421-nm, $\Delta g\approx0$ transition because $\gamma_{p}\ll\omega_{L}\ll\Gamma$ is satisfied throughout the majority of the Dy MOT's quadrupole field.  Specifically, Dy's optical pumping rate $\gamma_{p}\approx\bar{I}\Gamma/(1+\bar{I}+4\delta^2)$ is approximately equal to the light shift frequency, and the inequality is satisfied with the range of  values $[\gamma_{p},\omega_{L},\Gamma]=2\pi\cdot[0.8,1-28,32]$ MHz present in a typical Dy MOT with $\nabla B = 20$ G/cm and $0.04 < r < 1$ mm.  In addition, $\gamma_{p}/\omega_{L}<1$ for $\nabla\text{B}>10$ G/cm, which is close to the measured critical value $\nabla_{z}\text{B}_{c}/2=\nabla_{\rho}\text{B}_{c}\approx12$ G/cm.  We do not have a simple estimate for $\bar{I}_{c}=0.17$ and $I^{c}_{r}=1.4$, but we expect that 2D numerical simulations of the Dy MOT could explicate the origin of these specific values as well as the underlying mechanism behind the stripe orientation transitions.

\section{Conclusion}

In conclusion, we note the possibility that other highly magnetic lanthanides could exhibit anisotropic sub-Doppler cooling for large enough cooling beam intensity and field gradient:  This novel cooling mechanism should be observable in Er MOTs when larger MOT beam intensities are employed.  The highly magnetic atom Ho (9 $\mu{B}$, $\Delta g/g_{g}=1.7$\%)---of interest for quantum information processing~\cite{Saffman:2008}---would likely exhibit anisotropic sub-Doppler cooling as well, if it were proven amenable to repumper-less magneto-optical trapping.

\begin{acknowledgments}
We thank J{.} V{.} Porto and W{.} D{.} Phillips for discussions, J. Ye for a critical reading of the manuscript, and N. Burdick for technical assistance.  We acknowledge support from the NSF (PHY08-47469), AFOSR (FA9550-09-1-0079), and the Army Research Office MURI 
award W911NF0910406.
\end{acknowledgments}


\begin{thebibliography}{30}%
\makeatletter
\providecommand \@ifxundefined [1]{%
 \@ifx{#1\undefined}
}%
\providecommand \@ifnum [1]{%
 \ifnum #1\expandafter \@firstoftwo
 \else \expandafter \@secondoftwo
 \fi
}%
\providecommand \@ifx [1]{%
 \ifx #1\expandafter \@firstoftwo
 \else \expandafter \@secondoftwo
 \fi
}%
\providecommand \natexlab [1]{#1}%
\providecommand \enquote  [1]{``#1''}%
\providecommand \bibnamefont  [1]{#1}%
\providecommand \bibfnamefont [1]{#1}%
\providecommand \citenamefont [1]{#1}%
\providecommand \href@noop [0]{\@secondoftwo}%
\providecommand \href [0]{\begingroup \@sanitize@url \@href}%
\providecommand \@href[1]{\@@startlink{#1}\@@href}%
\providecommand \@@href[1]{\endgroup#1\@@endlink}%
\providecommand \@sanitize@url [0]{\catcode `\\12\catcode `\$12\catcode
  `\&12\catcode `\#12\catcode `\^12\catcode `\_12\catcode `\%12\relax}%
\providecommand \@@startlink[1]{}%
\providecommand \@@endlink[0]{}%
\providecommand \url  [0]{\begingroup\@sanitize@url \@url }%
\providecommand \@url [1]{\endgroup\@href {#1}{\urlprefix }}%
\providecommand \urlprefix  [0]{URL }%
\providecommand \Eprint [0]{\href }%
\providecommand \doibase [0]{http://dx.doi.org/}%
\providecommand \selectlanguage [0]{\@gobble}%
\providecommand \bibinfo  [0]{\@secondoftwo}%
\providecommand \bibfield  [0]{\@secondoftwo}%
\providecommand \translation [1]{[#1]}%
\providecommand \BibitemOpen [0]{}%
\providecommand \bibitemStop [0]{}%
\providecommand \bibitemNoStop [0]{.\EOS\space}%
\providecommand \EOS [0]{\spacefactor3000\relax}%
\providecommand \BibitemShut  [1]{\csname bibitem#1\endcsname}%
\let\auto@bib@innerbib\@empty
\bibitem [{\citenamefont {Fregoso}\ \emph {et~al.}(2009)\citenamefont
  {Fregoso}, \citenamefont {Sun}, \citenamefont {Fradkin},\ and\ \citenamefont
  {Lev}}]{Fregoso:2009}%
  \BibitemOpen
  \bibfield  {author} {\bibinfo {author} {\bibfnamefont {B.~M.}\ \bibnamefont
  {Fregoso}}, \bibinfo {author} {\bibfnamefont {K.}~\bibnamefont {Sun}},
  \bibinfo {author} {\bibfnamefont {E.}~\bibnamefont {Fradkin}}, \ and\
  \bibinfo {author} {\bibfnamefont {B.~L.}\ \bibnamefont {Lev}},\ }\href@noop
  {} {\bibfield  {journal} {\bibinfo  {journal} {New J. Phys.}\ }\textbf
  {\bibinfo {volume} {11}},\ \bibinfo {pages} {103003} (\bibinfo {year}
  {2009})}\BibitemShut {NoStop}%
\bibitem [{\citenamefont {Fregoso}\ and\ \citenamefont
  {Fradkin}(2009)}]{Fregoso:2009b}%
  \BibitemOpen
  \bibfield  {author} {\bibinfo {author} {\bibfnamefont {B.~M.}\ \bibnamefont
  {Fregoso}}\ and\ \bibinfo {author} {\bibfnamefont {E.}~\bibnamefont
  {Fradkin}},\ }\href@noop {} {\bibfield  {journal} {\bibinfo  {journal} {Phys.
  Rev. Lett.}\ }\textbf {\bibinfo {volume} {103}},\ \bibinfo {pages} {205301}
  (\bibinfo {year} {2009})}\BibitemShut {NoStop}%
\bibitem [{\citenamefont {Fregoso}\ and\ \citenamefont
  {Fradkin}(2010)}]{fregoso:2010}%
  \BibitemOpen
  \bibfield  {author} {\bibinfo {author} {\bibfnamefont {B.~M.}\ \bibnamefont
  {Fregoso}}\ and\ \bibinfo {author} {\bibfnamefont {E.}~\bibnamefont
  {Fradkin}},\ }\href@noop {} {\bibfield  {journal} {\bibinfo  {journal} {Phys.
  Rev. B}\ }\textbf {\bibinfo {volume} {81}},\ \bibinfo {pages} {214443}
  (\bibinfo {year} {2010})}\BibitemShut {NoStop}%
\bibitem [{\citenamefont {Lahaye}\ \emph {et~al.}(2009)\citenamefont {Lahaye},
  \citenamefont {Menotti}, \citenamefont {Santos}, \citenamefont {Lewenstein},\
  and\ \citenamefont {Pfau}}]{PfauReview09}%
  \BibitemOpen
  \bibfield  {author} {\bibinfo {author} {\bibfnamefont {T.}~\bibnamefont
  {Lahaye}}, \bibinfo {author} {\bibfnamefont {C.}~\bibnamefont {Menotti}},
  \bibinfo {author} {\bibfnamefont {L.}~\bibnamefont {Santos}}, \bibinfo
  {author} {\bibfnamefont {M.}~\bibnamefont {Lewenstein}}, \ and\ \bibinfo
  {author} {\bibfnamefont {T.}~\bibnamefont {Pfau}},\ }\href@noop {} {\bibfield
   {journal} {\bibinfo  {journal} {Rep. Prog. Phys.}\ }\textbf {\bibinfo
  {volume} {72}},\ \bibinfo {pages} {126401} (\bibinfo {year}
  {2009})}\BibitemShut {NoStop}%
\bibitem [{\citenamefont {Sogo}\ \emph {et~al.}(2009)\citenamefont {Sogo},
  \citenamefont {He}, \citenamefont {Miyakawa}, \citenamefont {Yi},
  \citenamefont {Lu},\ and\ \citenamefont {Pu}}]{Pu09}%
  \BibitemOpen
  \bibfield  {author} {\bibinfo {author} {\bibfnamefont {T.}~\bibnamefont
  {Sogo}}, \bibinfo {author} {\bibfnamefont {L.}~\bibnamefont {He}}, \bibinfo
  {author} {\bibfnamefont {T.}~\bibnamefont {Miyakawa}}, \bibinfo {author}
  {\bibfnamefont {S.}~\bibnamefont {Yi}}, \bibinfo {author} {\bibfnamefont
  {H.}~\bibnamefont {Lu}}, \ and\ \bibinfo {author} {\bibfnamefont
  {H.}~\bibnamefont {Pu}},\ }\href@noop {} {\bibfield  {journal} {\bibinfo
  {journal} {New J. Phys.}\ }\textbf {\bibinfo {volume} {11}},\ \bibinfo
  {pages} {055017} (\bibinfo {year} {2009})}\BibitemShut {NoStop}%
\bibitem [{\citenamefont {Ronen}\ and\ \citenamefont {Bohn}(2010)}]{Bohn10}%
  \BibitemOpen
  \bibfield  {author} {\bibinfo {author} {\bibfnamefont {S.}~\bibnamefont
  {Ronen}}\ and\ \bibinfo {author} {\bibfnamefont {J.~L.}\ \bibnamefont
  {Bohn}},\ }\href@noop {} {\bibfield  {journal} {\bibinfo  {journal} {Phys.
  Rev. A}\ }\textbf {\bibinfo {volume} {81}},\ \bibinfo {pages} {033601}
  (\bibinfo {year} {2010})}\BibitemShut {NoStop}%
\bibitem [{\citenamefont {{J. A. M. {Huhtam\"{a}ki}}}\ \emph
  {et~al.}(2010)\citenamefont {{J. A. M. {Huhtam\"{a}ki}}}, \citenamefont
  {Takahashi}, \citenamefont {{T. P. Simula}}, \citenamefont {Mizushima},\ and\
  \citenamefont {Machida}}]{Machida10}%
  \BibitemOpen
  \bibfield  {author} {\bibinfo {author} {\bibnamefont {{J. A. M.
  {Huhtam\"{a}ki}}}}, \bibinfo {author} {\bibfnamefont {M.}~\bibnamefont
  {Takahashi}}, \bibinfo {author} {\bibnamefont {{T. P. Simula}}}, \bibinfo
  {author} {\bibfnamefont {T.}~\bibnamefont {Mizushima}}, \ and\ \bibinfo
  {author} {\bibfnamefont {K.}~\bibnamefont {Machida}},\ }\href@noop {}
  {\bibfield  {journal} {\bibinfo  {journal} {Phys. Rev. A}\ }\textbf {\bibinfo
  {volume} {81}},\ \bibinfo {pages} {063623} (\bibinfo {year}
  {2010})}\BibitemShut {NoStop}%
\bibitem [{\citenamefont {Lu}\ \emph {et~al.}(2010)\citenamefont {Lu},
  \citenamefont {Youn},\ and\ \citenamefont {Lev}}]{Lu2010}%
  \BibitemOpen
  \bibfield  {author} {\bibinfo {author} {\bibfnamefont {M.}~\bibnamefont
  {Lu}}, \bibinfo {author} {\bibfnamefont {S.-H.}\ \bibnamefont {Youn}}, \ and\
  \bibinfo {author} {\bibfnamefont {B.~L.}\ \bibnamefont {Lev}},\ }\href@noop
  {} {\bibfield  {journal} {\bibinfo  {journal} {Phys. Rev. Lett.}\ }\textbf
  {\bibinfo {volume} {104}},\ \bibinfo {pages} {063001} (\bibinfo {year}
  {2010})}\BibitemShut {NoStop}%
\bibitem [{\citenamefont {Youn}\ \emph {et~al.}(2010)\citenamefont {Youn},
  \citenamefont {Lu}, \citenamefont {Ray},\ and\ \citenamefont
  {Lev}}]{Youn2010a}%
  \BibitemOpen
  \bibfield  {author} {\bibinfo {author} {\bibfnamefont {S.-H.}\ \bibnamefont
  {Youn}}, \bibinfo {author} {\bibfnamefont {M.}~\bibnamefont {Lu}}, \bibinfo
  {author} {\bibfnamefont {U.}~\bibnamefont {Ray}}, \ and\ \bibinfo {author}
  {\bibfnamefont {B.~L.}\ \bibnamefont {Lev}},\ }\href@noop {} {} (\bibinfo
  {year} {2010}),\ \Eprint {http://arxiv.org/abs/1007.1480} {arXiv:1007.1480}
  \BibitemShut {NoStop}%
\bibitem [{\citenamefont {Yi}\ \emph {et~al.}(2007)\citenamefont {Yi},
  \citenamefont {Li},\ and\ \citenamefont {Sun}}]{Yi:2007}%
  \BibitemOpen
  \bibfield  {author} {\bibinfo {author} {\bibfnamefont {S.}~\bibnamefont
  {Yi}}, \bibinfo {author} {\bibfnamefont {T.}~\bibnamefont {Li}}, \ and\
  \bibinfo {author} {\bibfnamefont {C.~P.}\ \bibnamefont {Sun}},\ }\href@noop
  {} {\bibfield  {journal} {\bibinfo  {journal} {Phys. Rev. Lett.}\ }\textbf
  {\bibinfo {volume} {98}},\ \bibinfo {pages} {260405} (\bibinfo {year}
  {2007})}\BibitemShut {NoStop}%
\bibitem [{\citenamefont {Dalibard}\ and\ \citenamefont
  {Cohen-Tannoudji}(1989)}]{Dalibard89}%
  \BibitemOpen
  \bibfield  {author} {\bibinfo {author} {\bibfnamefont {J.}~\bibnamefont
  {Dalibard}}\ and\ \bibinfo {author} {\bibfnamefont {C.}~\bibnamefont
  {Cohen-Tannoudji}},\ }\href@noop {} {\bibfield  {journal} {\bibinfo
  {journal} {J. Opt. Soc. Am. B}\ }\textbf {\bibinfo {volume} {6}},\ \bibinfo
  {pages} {2023} (\bibinfo {year} {1989})}\BibitemShut {NoStop}%
\bibitem [{\citenamefont {Ungar}\ \emph {et~al.}(1989)\citenamefont {Ungar},
  \citenamefont {Weiss}, \citenamefont {Riis},\ and\ \citenamefont
  {Chu}}]{Chu89}%
  \BibitemOpen
  \bibfield  {author} {\bibinfo {author} {\bibfnamefont {P.}~\bibnamefont
  {Ungar}}, \bibinfo {author} {\bibfnamefont {D.~S.}\ \bibnamefont {Weiss}},
  \bibinfo {author} {\bibfnamefont {E.}~\bibnamefont {Riis}}, \ and\ \bibinfo
  {author} {\bibfnamefont {S.}~\bibnamefont {Chu}},\ }\href@noop {} {\bibfield
  {journal} {\bibinfo  {journal} {J. Opt. Soc. Am. B}\ }\textbf {\bibinfo
  {volume} {6}},\ \bibinfo {pages} {2058} (\bibinfo {year} {1989})}\BibitemShut
  {NoStop}%
\bibitem [{\citenamefont {Metcalf}\ and\ \citenamefont {{P. van der
  Straten}}(1999)}]{MetcalfBook99}%
  \BibitemOpen
  \bibfield  {author} {\bibinfo {author} {\bibfnamefont {H.~J.}\ \bibnamefont
  {Metcalf}}\ and\ \bibinfo {author} {\bibnamefont {{P. van der Straten}}},\
  }\href@noop {} {\emph {\bibinfo {title} {Laser Cooling and Trapping}}}\
  (\bibinfo  {publisher} {Springer-Verlag, New York},\ \bibinfo {year}
  {1999})\BibitemShut {NoStop}%
\bibitem [{\citenamefont {Drewsen}\ \emph {et~al.}(1994)\citenamefont
  {Drewsen}, \citenamefont {Laurent}, \citenamefont {Nadir}, \citenamefont
  {Santarelli}, \citenamefont {Clairon}, \citenamefont {Castin}, \citenamefont
  {Grison},\ and\ \citenamefont {Salomon}}]{Salomon94}%
  \BibitemOpen
  \bibfield  {author} {\bibinfo {author} {\bibfnamefont {M.}~\bibnamefont
  {Drewsen}}, \bibinfo {author} {\bibfnamefont {P.}~\bibnamefont {Laurent}},
  \bibinfo {author} {\bibfnamefont {A.}~\bibnamefont {Nadir}}, \bibinfo
  {author} {\bibfnamefont {G.}~\bibnamefont {Santarelli}}, \bibinfo {author}
  {\bibfnamefont {A.}~\bibnamefont {Clairon}}, \bibinfo {author} {\bibfnamefont
  {Y.}~\bibnamefont {Castin}}, \bibinfo {author} {\bibfnamefont
  {D.}~\bibnamefont {Grison}}, \ and\ \bibinfo {author} {\bibfnamefont
  {C.}~\bibnamefont {Salomon}},\ }\href@noop {} {\bibfield  {journal} {\bibinfo
   {journal} {Appl. Phys. B}\ }\textbf {\bibinfo {volume} {59}},\ \bibinfo
  {pages} {283} (\bibinfo {year} {1994})}\BibitemShut {NoStop}%
\bibitem [{\citenamefont {Walhout}\ \emph {et~al.}(1992)\citenamefont
  {Walhout}, \citenamefont {Dalibard}, \citenamefont {Rolston},\ and\
  \citenamefont {Phillips}}]{Phillips92}%
  \BibitemOpen
  \bibfield  {author} {\bibinfo {author} {\bibfnamefont {M.}~\bibnamefont
  {Walhout}}, \bibinfo {author} {\bibfnamefont {J.}~\bibnamefont {Dalibard}},
  \bibinfo {author} {\bibfnamefont {S.~L.}\ \bibnamefont {Rolston}}, \ and\
  \bibinfo {author} {\bibfnamefont {W.~D.}\ \bibnamefont {Phillips}},\
  }\href@noop {} {\bibfield  {journal} {\bibinfo  {journal} {J. Opt. Soc. Am.
  B}\ }\textbf {\bibinfo {volume} {9}},\ \bibinfo {pages} {1997} (\bibinfo
  {year} {1992})}\BibitemShut {NoStop}%
\bibitem [{\citenamefont {Werner}\ \emph {et~al.}(1992)\citenamefont {Werner},
  \citenamefont {Wallis},\ and\ \citenamefont {Ertmer}}]{Ertmer92}%
  \BibitemOpen
  \bibfield  {author} {\bibinfo {author} {\bibfnamefont {J.}~\bibnamefont
  {Werner}}, \bibinfo {author} {\bibfnamefont {H.}~\bibnamefont {Wallis}}, \
  and\ \bibinfo {author} {\bibfnamefont {W.}~\bibnamefont {Ertmer}},\
  }\href@noop {} {\bibfield  {journal} {\bibinfo  {journal} {Opt. Commun.}\
  }\textbf {\bibinfo {volume} {94}},\ \bibinfo {pages} {525} (\bibinfo {year}
  {1992})}\BibitemShut {NoStop}%
\bibitem [{Note1()}]{Note1}%
  \BibitemOpen
  \bibinfo {note} {The $^{87}$Sr system, which has a negligible $g_{g}$, also
  exhibits intra-MOT sub-Doppler cooling~\cite {YeSr03}.}\BibitemShut {Stop}%
\bibitem [{\citenamefont {McClelland}\ and\ \citenamefont
  {Hanssen}(2006)}]{Mcclelland:2006}%
  \BibitemOpen
  \bibfield  {author} {\bibinfo {author} {\bibfnamefont {J.~J.}\ \bibnamefont
  {McClelland}}\ and\ \bibinfo {author} {\bibfnamefont {J.~L.}\ \bibnamefont
  {Hanssen}},\ }\href@noop {} {\bibfield  {journal} {\bibinfo  {journal} {Phys.
  Rev. Lett.}\ }\textbf {\bibinfo {volume} {96}},\ \bibinfo {pages} {143005}
  (\bibinfo {year} {2006})}\BibitemShut {NoStop}%
\bibitem [{\citenamefont {Martin}\ \emph {et~al.}(1978)\citenamefont {Martin},
  \citenamefont {Zalubas},\ and\ \citenamefont {Hagan}}]{Martin:1978}%
  \BibitemOpen
  \bibfield  {author} {\bibinfo {author} {\bibfnamefont {W.~C.}\ \bibnamefont
  {Martin}}, \bibinfo {author} {\bibfnamefont {R.}~\bibnamefont {Zalubas}}, \
  and\ \bibinfo {author} {\bibfnamefont {L.}~\bibnamefont {Hagan}},\
  }\href@noop {} {\emph {\bibinfo {title} {Atomic Energy Levels--The Rare Earth
  Elements}}}\ (\bibinfo  {publisher} {NSRDS-NBS, \textbf{60}},\ \bibinfo
  {address} {Washington, D.C.},\ \bibinfo {year} {1978})\BibitemShut {NoStop}%
\bibitem [{\citenamefont {Berglund}\ \emph {et~al.}(2007)\citenamefont
  {Berglund}, \citenamefont {Lee},\ and\ \citenamefont
  {McClelland}}]{Berglund:2007}%
  \BibitemOpen
  \bibfield  {author} {\bibinfo {author} {\bibfnamefont {A.~J.}\ \bibnamefont
  {Berglund}}, \bibinfo {author} {\bibfnamefont {S.~A.}\ \bibnamefont {Lee}}, \
  and\ \bibinfo {author} {\bibfnamefont {J.~J.}\ \bibnamefont {McClelland}},\
  }\href@noop {} {\bibfield  {journal} {\bibinfo  {journal} {Phys. Rev. A}\
  }\textbf {\bibinfo {volume} {76}},\ \bibinfo {pages} {053418} (\bibinfo
  {year} {2007})}\BibitemShut {NoStop}%
\bibitem [{\citenamefont {Xu}\ \emph {et~al.}(2003)\citenamefont {Xu},
  \citenamefont {Loftus}, \citenamefont {Dunn}, \citenamefont {Greene},
  \citenamefont {Hall}, \citenamefont {Gallagher},\ and\ \citenamefont
  {Ye}}]{YeSr03}%
  \BibitemOpen
  \bibfield  {author} {\bibinfo {author} {\bibfnamefont {X.}~\bibnamefont
  {Xu}}, \bibinfo {author} {\bibfnamefont {T.~H.}\ \bibnamefont {Loftus}},
  \bibinfo {author} {\bibfnamefont {J.~W.}\ \bibnamefont {Dunn}}, \bibinfo
  {author} {\bibfnamefont {C.~H.}\ \bibnamefont {Greene}}, \bibinfo {author}
  {\bibfnamefont {J.~L.}\ \bibnamefont {Hall}}, \bibinfo {author}
  {\bibfnamefont {A.}~\bibnamefont {Gallagher}}, \ and\ \bibinfo {author}
  {\bibfnamefont {J.}~\bibnamefont {Ye}},\ }\href@noop {} {\bibfield  {journal}
  {\bibinfo  {journal} {Phys. Rev. Lett.}\ }\textbf {\bibinfo {volume} {90}},\
  \bibinfo {pages} {193002} (\bibinfo {year} {2003})}\BibitemShut {NoStop}%
\bibitem [{\citenamefont {Shang}\ \emph {et~al.}(1991)\citenamefont {Shang},
  \citenamefont {Sheehy}, \citenamefont {Metcalf}, \citenamefont {van~der
  Straten},\ and\ \citenamefont {Nienhuis}}]{Metcalf91}%
  \BibitemOpen
  \bibfield  {author} {\bibinfo {author} {\bibfnamefont {S.-Q.}\ \bibnamefont
  {Shang}}, \bibinfo {author} {\bibfnamefont {B.}~\bibnamefont {Sheehy}},
  \bibinfo {author} {\bibfnamefont {H.}~\bibnamefont {Metcalf}}, \bibinfo
  {author} {\bibfnamefont {P.}~\bibnamefont {van~der Straten}}, \ and\ \bibinfo
  {author} {\bibfnamefont {G.}~\bibnamefont {Nienhuis}},\ }\href@noop {}
  {\bibfield  {journal} {\bibinfo  {journal} {Phys. Rev. Lett.}\ }\textbf
  {\bibinfo {volume} {67}},\ \bibinfo {pages} {1094} (\bibinfo {year}
  {1991})}\BibitemShut {NoStop}%
\bibitem [{\citenamefont {van~der Straten}\ \emph {et~al.}(1993)\citenamefont
  {van~der Straten}, \citenamefont {Shang}, \citenamefont {Sheehy},
  \citenamefont {Metcalf},\ and\ \citenamefont {Nienhuis}}]{vanderStraten93}%
  \BibitemOpen
  \bibfield  {author} {\bibinfo {author} {\bibfnamefont {P.}~\bibnamefont
  {van~der Straten}}, \bibinfo {author} {\bibfnamefont {S.-Q.}\ \bibnamefont
  {Shang}}, \bibinfo {author} {\bibfnamefont {B.}~\bibnamefont {Sheehy}},
  \bibinfo {author} {\bibfnamefont {H.}~\bibnamefont {Metcalf}}, \ and\
  \bibinfo {author} {\bibfnamefont {G.}~\bibnamefont {Nienhuis}},\ }\href@noop
  {} {\bibfield  {journal} {\bibinfo  {journal} {Phys. Rev. A}\ }\textbf
  {\bibinfo {volume} {47}},\ \bibinfo {pages} {4160} (\bibinfo {year}
  {1993})}\BibitemShut {NoStop}%
\bibitem [{\citenamefont {Chang}\ \emph {et~al.}(2002)\citenamefont {Chang},
  \citenamefont {Kwon}, \citenamefont {Lee},\ and\ \citenamefont
  {Minogin}}]{Lee02}%
  \BibitemOpen
  \bibfield  {author} {\bibinfo {author} {\bibfnamefont {S.}~\bibnamefont
  {Chang}}, \bibinfo {author} {\bibfnamefont {T.-Y.}\ \bibnamefont {Kwon}},
  \bibinfo {author} {\bibfnamefont {H.-S.}\ \bibnamefont {Lee}}, \ and\
  \bibinfo {author} {\bibfnamefont {V.~G.}\ \bibnamefont {Minogin}},\
  }\href@noop {} {\bibfield  {journal} {\bibinfo  {journal} {Phys. Rev. A}\
  }\textbf {\bibinfo {volume} {66}},\ \bibinfo {pages} {043404} (\bibinfo
  {year} {2002})}\BibitemShut {NoStop}%
\bibitem [{Note2()}]{Note2}%
  \BibitemOpen
  \bibinfo {note} {Boundaries between the striped MOT regimes are blurred with
  substantial beam misalignment.}\BibitemShut {Stop}%
\bibitem [{\citenamefont {Kim}\ \emph {et~al.}(2004)\citenamefont {Kim},
  \citenamefont {Noh}, \citenamefont {Ha},\ and\ \citenamefont
  {Jhe}}]{Jhe:2004}%
  \BibitemOpen
  \bibfield  {author} {\bibinfo {author} {\bibfnamefont {K.}~\bibnamefont
  {Kim}}, \bibinfo {author} {\bibfnamefont {H.-R.}\ \bibnamefont {Noh}},
  \bibinfo {author} {\bibfnamefont {H.-J.}\ \bibnamefont {Ha}}, \ and\ \bibinfo
  {author} {\bibfnamefont {W.}~\bibnamefont {Jhe}},\ }\href@noop {} {\bibfield
  {journal} {\bibinfo  {journal} {Phys. Rev. A}\ }\textbf {\bibinfo {volume}
  {69}},\ \bibinfo {pages} {33406} (\bibinfo {year} {2004})}\BibitemShut
  {NoStop}%
\bibitem [{Note3()}]{Note3}%
  \BibitemOpen
  \bibinfo {note} {Careful characterization of minority population and
  temperature for MOTs with $I_{z}/I_{\rho }>1$ was inhibited by increased MOT
  population instability in these regimes.}\BibitemShut {Stop}%
\bibitem [{Note4()}]{Note4}%
  \BibitemOpen
  \bibinfo {note} {We observe, by imaging in the $\rho $-plane, that the oblate
  stripe is indeed azimuthally symmetric.}\BibitemShut {Stop}%
\bibitem [{Note5()}]{Note5}%
  \BibitemOpen
  \bibinfo {note} {A review of such a treatment is beyond the scope of the
  current work. However, Refs.~\cite {vanderStraten93,Lee02} provide detailed
  analytical and numerical calculations of the force and diffusion felt by
  atoms in various relative orientations of (a large) magnetic field and
  $\sigma ^{+}$-$\sigma ^{-}$ light; see Figs.~4 and 5 and Table 1 of
  Ref.~\cite {vanderStraten93} and Fig.~4 of~\cite {Lee02} for force versus
  velocity plots and additional details of the VSR cases invoked in Sec.~\ref
  {VSR}}\BibitemShut {NoStop}%
\bibitem [{\citenamefont {Saffman}\ and\ \citenamefont
  {M{\o}lmer}(2008)}]{Saffman:2008}%
  \BibitemOpen
  \bibfield  {author} {\bibinfo {author} {\bibfnamefont {M.}~\bibnamefont
  {Saffman}}\ and\ \bibinfo {author} {\bibfnamefont {K.}~\bibnamefont
  {M{\o}lmer}},\ }\href@noop {} {\bibfield  {journal} {\bibinfo  {journal}
  {Phys. Rev. A}\ }\textbf {\bibinfo {volume} {78}},\ \bibinfo {pages} {012336}
  (\bibinfo {year} {2008})}\BibitemShut {NoStop}%
\end{thebibliography}
%

\end{document}